\documentclass[traditabstract]{aa}
\usepackage{txfonts}
\usepackage{epsfig}
\usepackage{graphicx}
\usepackage{natbib}
\usepackage{url}
\usepackage{twoopt}
\usepackage[breaklinks=true]{hyperref}
\usepackage{dcolumn}

\usepackage{color}
\usepackage{ulem}

\begin{document}

\title{New study of the  line profiles of sodium perturbed by H$_2$}
\author{N.~F.~Allard    \inst{1,2} 
   \and F. Spiegelman  \inst{3} 
   \and T.~Leininger    \inst{3}
   \and   P. Molliere   \inst{4} 
 }

\offprints{nicole.allard@obspm.fr}

\institute{
GEPI, Observatoire de Paris,  PSL Research University, UMR 8111, CNRS, 
Sorbonne Paris Cit\'e,
 61, Avenue de l'Observatoire, F-75014 Paris, France \\
\email{nicole.allard@obspm.fr}
\and
Institut d'Astrophysique de Paris,  UMR7095, CNRS, Universit\'e Paris VI,
 98bis Boulevard Arago, PARIS, France\\
\and
Laboratoire de Physique et Chimie Quantique, Universit\'e de      
           Toulouse (UPS) and CNRS, 118 route de Narbonne, 
           F-31400 Toulouse,   France \\
          \and
Leiden Observatory, Leiden University, Postbus 9513, 2300 RA Leiden, The Netherlands            \\
}

\date{march29 ...; accepted ...}

\abstract{
The opacity of alkali atoms, most importantly of Na and K, plays a
crucial role in the atmospheres of brown dwarfs and exoplanets.
  We present a comprehensive study  of Na--H$_2$ collisional
  profiles at temperatures  from 500~ to 3000~K, the temperatures
  prevailing in the atmosphere of brown dwarfs and Jupiter-mass planets.
  The relevant H$_2$ perturber densities reach several 10$^{19}$~cm$^{-3}$
  in hot ($T_{\rm eff}\gtrsim 1500$ K) Jupiter-mass planets and can exceed
  10$^{20}$~cm$^{-3}$  for more massive or cooler objects. Accurate pressure-broadened profiles that are valid at high densities of H$_2$ should be
  incorporated into spectral models.
  Unified profiles of  sodium  perturbed by molecular hydrogen were calculated in the semi-classical approach 
   using  up-to-date molecular data.
New Na--H$_2$ collisional profiles and their effects on the synthetic spectra 
of  brown dwarfs  and hot Jupiters computed
with {\it petitCODE} are presented.
}


\authorrunning{N.~F.~Allard et al.}
\titlerunning{New study of the  line profiles of sodium perturbed by H$_2$}

\maketitle

\section{Introduction}
        Alkali atoms are an important class of absorbers for modeling and
understanding the spectra of self-luminous objects such
as brown dwarfs and directly imaged planets.
The wings of the sodium and potassium resonance lines
in the optical are particularly important because they serve as a source of
pseudo-continuum opacity, reaching into the near-infrared (NIR) wavelengths in
the case of potassium. The relevance of these alkali species for the
atmospheres of such self-luminous objects is studied and discussed in
detail in \citet{burrows2001}. Moreover, it has been shown that the exact
shape of the wings of the alkali lines, especially the red wings of the
K doublet, affects the atmospheric structure, and that different
treatments of the line wings can lead to differences in the temperature
profiles at larger pressures \citep[see, e.g.,][]{baudinomolliere2017}.

 The class of transiting exoplanets (especially the so-called hot
  Jupiters) is also affected by the presence of the alkalis. Key observations
  with space- and ground-based telescopes have shed light on the conditions
  and composition of their atmospheres. Sodium was first detected in the
  atmosphere of HD209458b  \citep{charbonneau2002}, and
is now routinely detected in many hot Jupiters from the ground and
from space at high and low resolution.
For examples, see \citet{snellen2008}, and the compilation of spectra
in \citet{sing2016} and  \citet{pino2018}.
Recently, the wing absorption of sodium was probed from the ground with
the FORS2 observations by \citet{Nikolov2018}.

 For irradiated gas planets
of intermediate temperature, the absorption of stellar light by the Na and
K doublet line wings in the optical represents an important heating
source \citep[e.g.,][]{mollierevanboekel2015}.
Moreover, in the absence of strong cloud absorption, Na and K are the
only significant absorbers of the flux of the host star in the optical.
For higher planetary temperatures (T $\gtrsim$ 2000 K), additional
 absorption by metal oxides, hydrides, atoms, and ions can become
  important \citep[see, e.g.,][]{arcangelidesert2018,lothringerbarman2018}
  and \citep{lothringerbarman2019}.
Coupling of Na and K abundance to energy transfer causes the atmospheric temperature structures of hot Jupiters to be very sensitive to the shapes of the Na and K doublet lines.
For special chemical conditions
where the planetary cooling opacity is low, the heating by alkali atoms
can even create inversions \citep{mollierevanboekel2015}.
For self-luminous planets, the alkali opacities are important by virtue of
their line wings as well. The red wings of the K doublet line
in particular can control the flux escaping the planets in the Y band, which can also
affect the planetary structure \citep{baudinomolliere2017}.

In continuation with \citet{allard2016b} we present new unified 
line profiles   of  neutral Na perturbed by H$_2$ using {\it ab initio}
Na--H$_2$  potentials and  transition dipole moments.
In \citet{allard2003}
we  presented  absorption  profiles of  sodium  perturbed  by  molecular
hydrogen. The line profiles were calculated in  a unified line shape
semi-classical theory  \citep{allard1999} using \citet{rossi1985}
pseudo-potentials.
Reliable calculations of pressure broadening in the far spectral wings
require accurate potential-energy curves describing the interaction of
the ground and excited states of Na with H$_2$. 
In~\cite{allard2003} and \cite{tinetti2007}
we  presented the first applications of semi-classical
  profiles of  sodium and
potassium perturbed  by   molecular  hydrogen
to the modeling of brown dwarfs and extra-solar planets.
The 3$s$-3$p$ resonance line profiles were calculated 
using the \cite{rossi1985}  pseudo-potentials (hereafter labeled RP85).
{\it Ab initio}\/ calculations of the potentials (hereafter labeled FS11) of 
Na--H$_2$  were  computed by one of us (FS) and compared to
pseudo-potentials of \citet{rossi1985} in \citet{allard2012b}.
We  highlighted the regions of interest near the Na--H$_2$ quasi-molecular 
satellites  for comparison with previous results described  
in \citet{allard2003}.  We also
compared with  laboratory absorption spectra.
We  extended this work to other excited states to allow a comprehensive
determination of the spectrum \citep{allard2012a}.
Nevertheless, tables of Na--H$_2$ absorption coefficients which are currently
used for the construction of model atmospheres and synthetic
spectra  which have been generated from the line profiles
reported in \cite{allard2003} needed to be up to date.

We have now  extended the calculations of the Na--H$_2$ 
potential energy surfaces to the 5$s$ state, and improved the accuracy for lower states.
The new {\it  ab initio} calculations of the potentials 
 are carried out for the C$_{2v}$  (T-shape) 
 symmetry group and the  C$_{\infty v}$ (linear) symmetry group.
 In this paper we restrict  our study 
 to the resonance 3$s$--3$p$ line and will use the 
 potentials of the more excited
  states described in Sect.~\ref{sec:pot} for a subsequent paper devoted to the
  line profiles of the sodium lines in the NIR.
In addition,  the transition dipole
moments for the resonance line absorption, as a function of the geometry
of the Na--H$_2$ system are  presented.
The improvement over our previous work \citep{allard2012b} 
consists in a better determination of the
 long-range part of the Na--H$_2$ potential curves. The inclusion 
 of the spin-orbit coupling together with this improvement allow  the
  determination of the individual line widths of the two components of the resonance
  3$s$-3$p$ doublet (Sect.~\ref{sec:temperature}). 
   We  illustrate the evolution of  the
absorption spectra of Na--H$_2$ collisional  profiles 
 for the densities  and temperatures
prevailing in the atmospheres of cool brown dwarf stars and extrasolar planets.
The  new opacity tables of Na--H$_2$  have  been incorporated
into atmosphere calculations of  self-luminous planets and  hot Jupiters.
 The atmospheric models presented in  Sect.~\ref{sec:astro}
  have been calculated with {\it petitCODE},
a well-tested code that solves for the 1D structures
of exoplanet atmospheres in radiative-convective and thermochemical
equilibrium. Gas and optionally cloud opacities can be included, and
scattering is treated in the structure and spectral calculations.
{\it  petitCODE} is described in
\citet{mollierevanboekel2015,mollierevanboekel2016}.

\begin{figure}
\centering
\vspace{8mm}
\includegraphics[width=12cm,angle=-90]{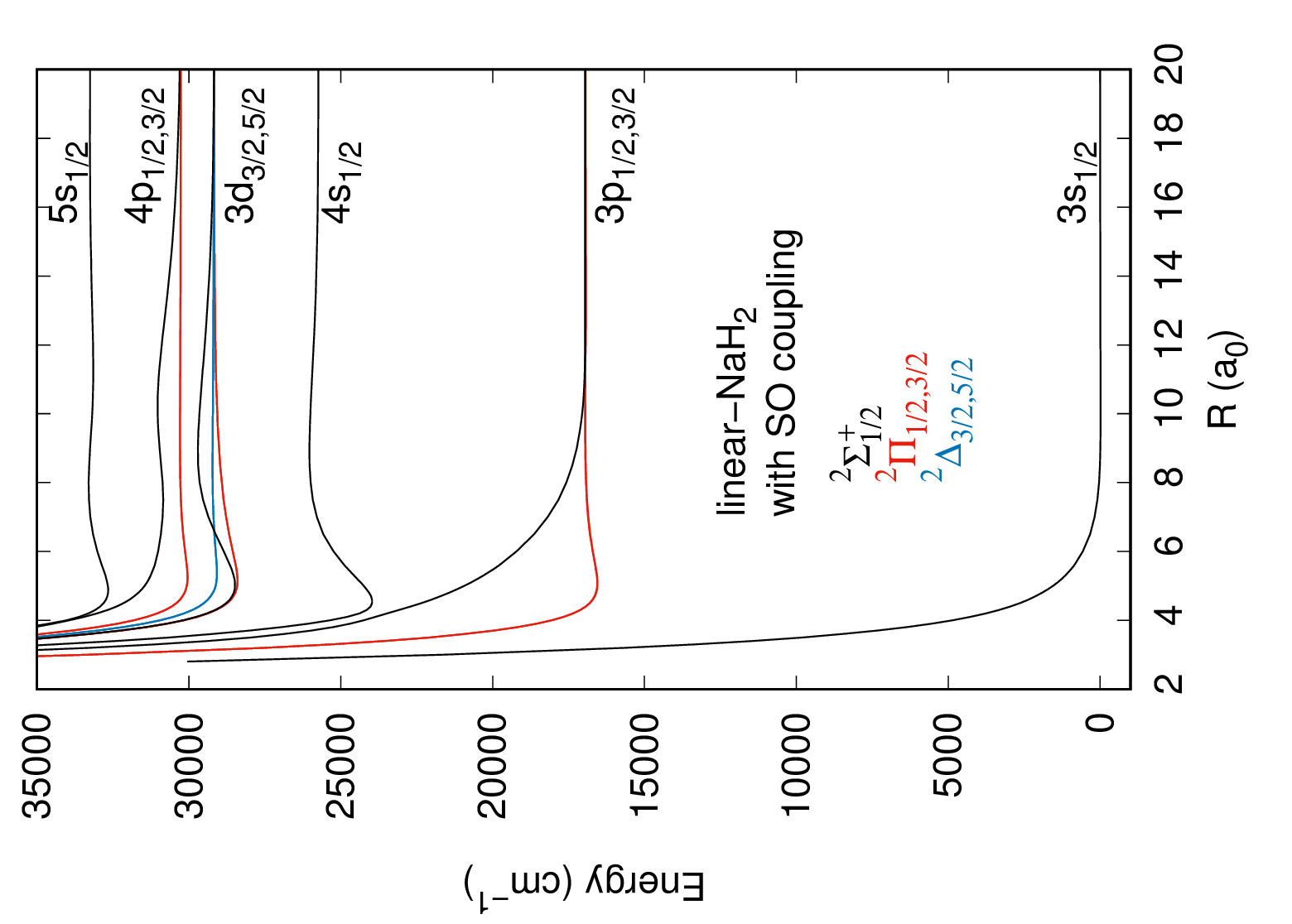}
\includegraphics[width=12cm,angle=-90]{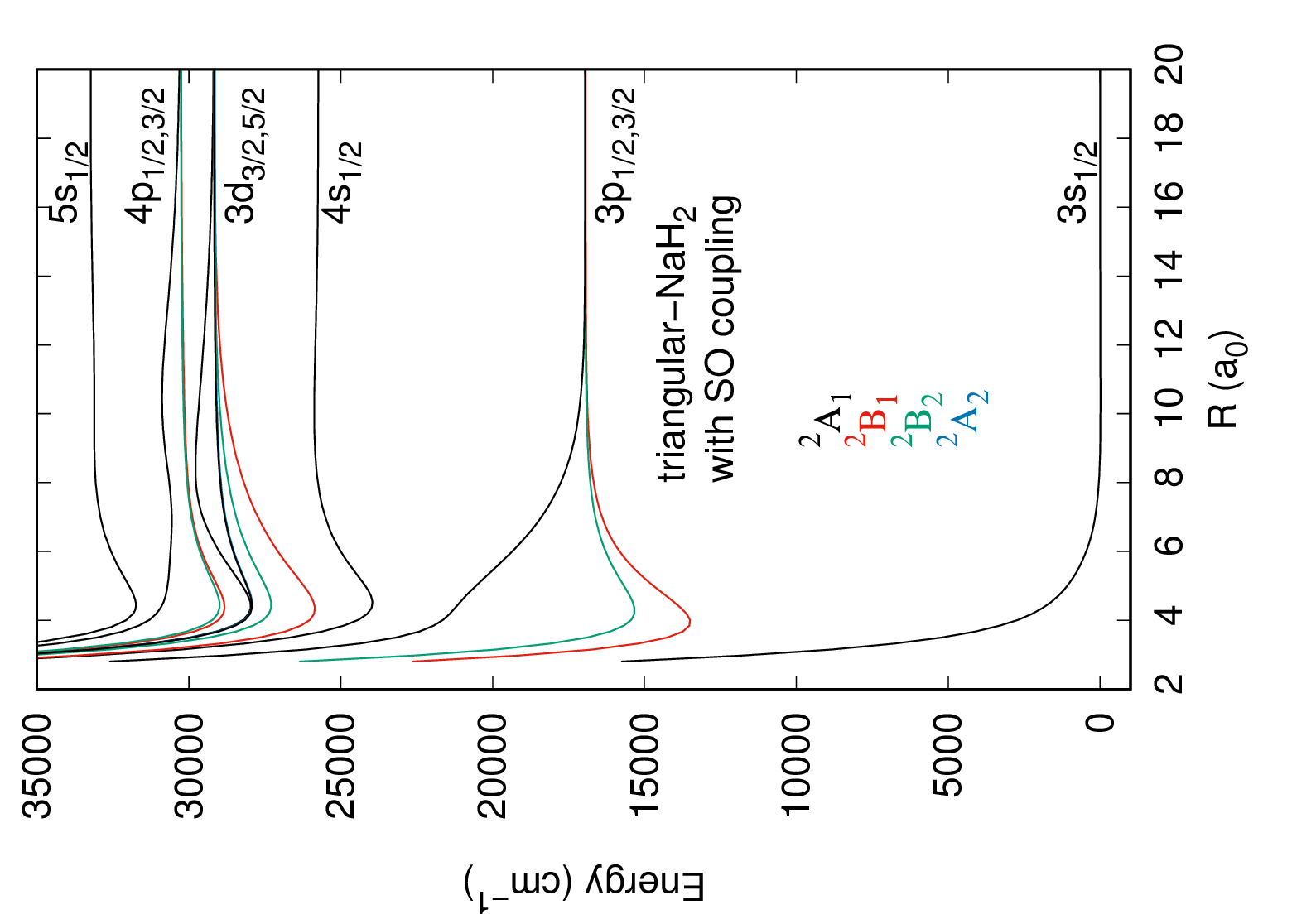}
\caption{Potential curves of  the
  Na--H$_2$ molecule for the  C$_{\infty v}$  (top) and $C_{2v}$
  symmetries (bottom).
  For the $C_{2v}$ case, the symmetry labeling corresponds to the
  convention of the reference plane  as that containing the molecule
  and may be  different from that of previous publications.
  We note that states 1$^2A_2$ and 4$^2A_1$ correlated with the
  $3d$ asymptote are superimposed at the scale of the figure.} 
\label{fig:potot}
\end{figure}

\begin{figure}
\centering
\includegraphics[width=6cm,angle=-90]{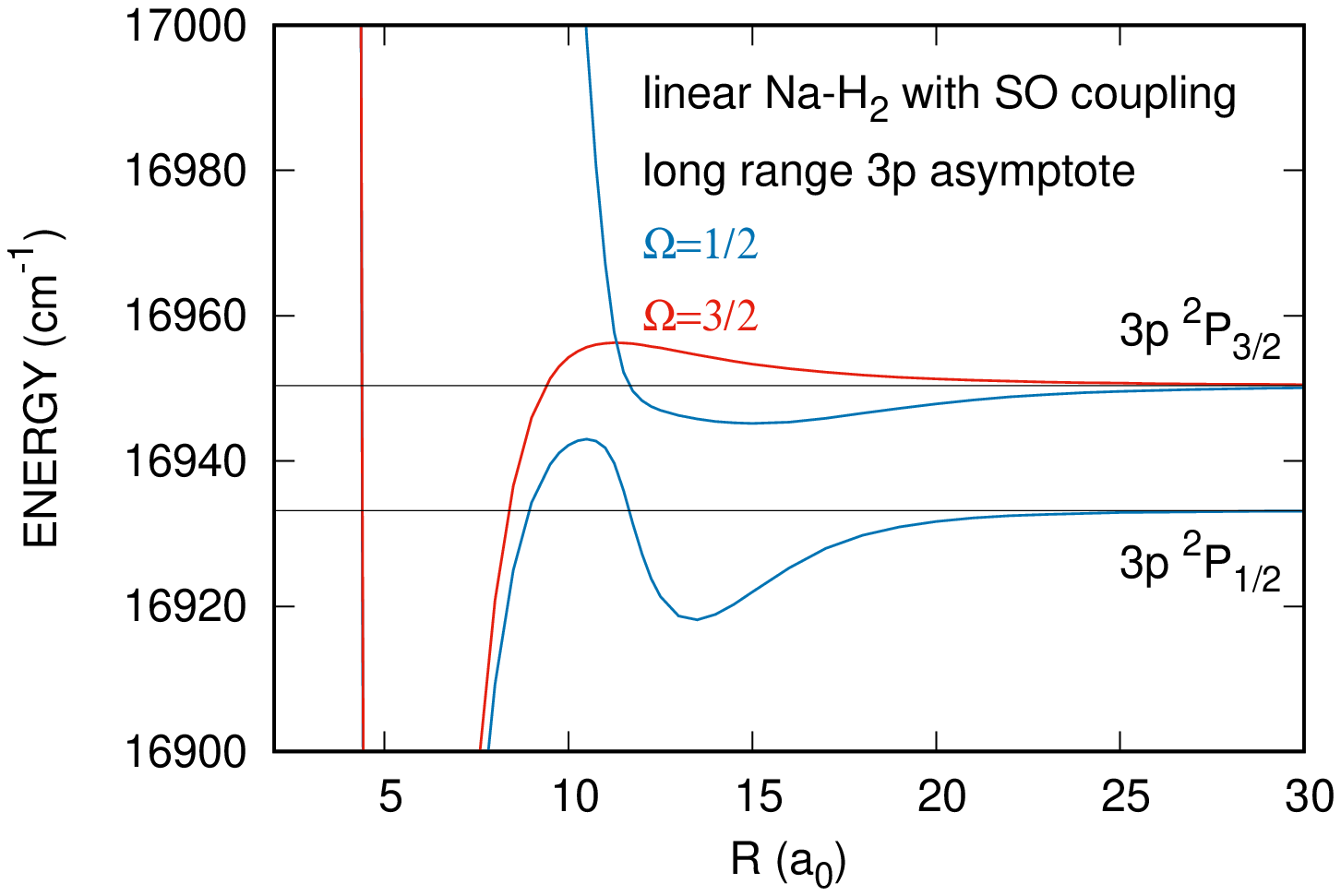}
\caption{Long-range potential curves of NaH$_2$ correlated with
  the $3p_{1/2}$ and $3p_{3/2}$ asymptotes in $C_{\infty v}$ symmetry.}  
\label{fig:LR3pl}
\end{figure}

\begin{figure}
\centering
\includegraphics[width=6cm,angle=-90]{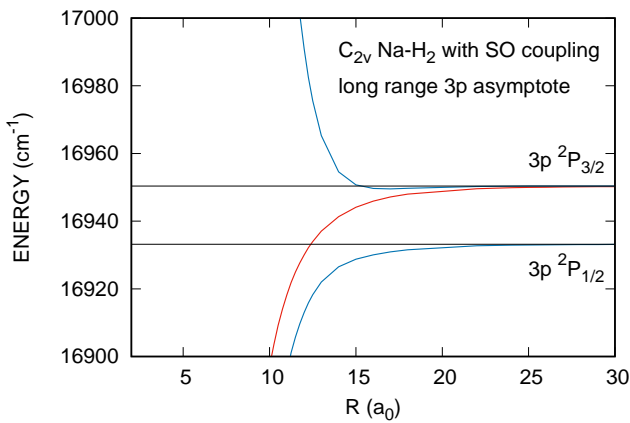}
\caption{Long-range potential curves of NaH$_2$ correlated with
  the $3p_{1/2}$ and $3p_{3/2}$ asymptotes in $C_{2v}$ symmetry.}  
\label{fig:LR3pt}
\end{figure}

\section{Na--H$_2$ potentials including
spin-orbit coupling \label{sec:pot}}

The  {\it ab initio} calculations of the potentials (hereafter labeled S17) 
 were carried out for the C$_{2v}$  (T-shape) 
symmetry group and the  C$_{\infty v}$ (linear) symmetry group in a 
wide range of distances $R$ between the Na atom and the center-of-mass of
the  H$_2$ molecule.
The  potentials were calculated with the MOLPRO package \citep{molpro2012}
and are shown in Figs.~\ref{fig:potot}-\ref{fig:LR3pt}.
In the calculation of the complex, the  bond length of H$_2$ 
was kept fixed at the equilibrium value $r_e$~=~1.401~a.u.
and the approach is along the $z$ coordinate axis. 
As in our previous calculations on  KH$_2$ 
\citep{allard2007a}  and NaH$_2$ \citep{allard2012b}, we  
used a single active electron description
of the sodium atom complemented with a core polarisation operator
to include the core response. The effective core potential is the ECP10SDF
effective core potential of the Stuttgart
group \citep{nicklass1995}.
  The core-polarization  uses the formulation of  \citet{muller1984a}
with the parameters $\alpha$=0.997 $a_0^3$ and $\rho_c=0.62$ 
for core polarizability and the cut-off parameter of the CPP
operator, respectively (corresponding  to the smooth cut-off  expression
defined in MOLPRO).
We used relatively extensive Gaussian-type basis sets (GTOs) to
describe the three active electrons, namely  8$s$5$p$6$d$8$f$4$g$ for Na
and an spdf AV5Z basis set for each hydrogen atom.
With this basis set, the one-electron scheme for sodium  describes
the transitions to the 
excited states of Na  with an accuracy better than 25 cm$^{-1}$ up to 5$s$, as 
illustrated in Table~1
and provides a fair account of
correlation  for the electrons of H$_2$.

\begin{figure}
\centering
\includegraphics[width=6cm,angle=-90]{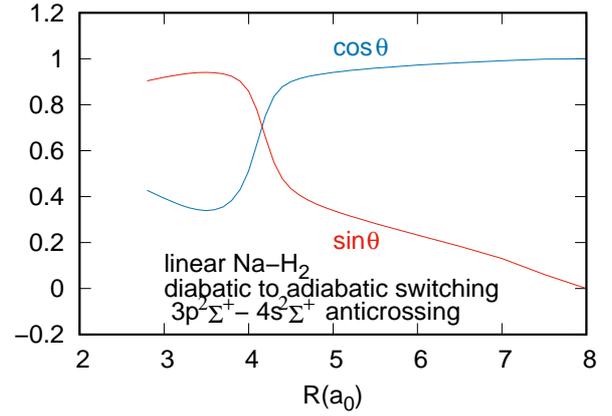}
\includegraphics[width=6cm,angle=-90]{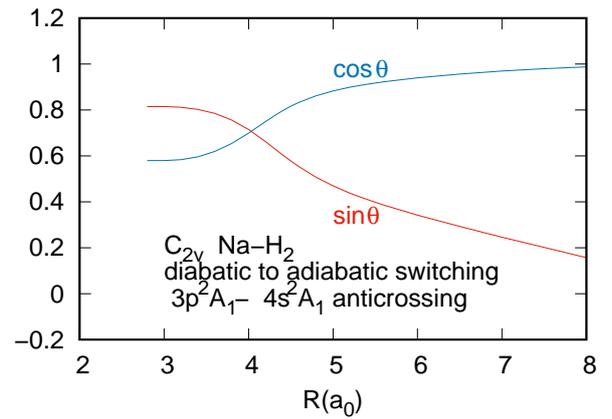}
\caption{Coefficients ($cos\theta$ and $sin\theta$)) of
  the 3$p$/4$p$ diabatization 
 in $C_{\infty v}$ (top) and $C_{2v}$ symmetries (bottom).} 
\label{fig:diab}
\end{figure}

The determination of the electronic structure of NaH$_2$ was
carried  out
using the Multi Reference Configuration Interaction (MRCI) scheme
of the MOLPRO package using the orbitals of NaH$_2^+$,  which should provide
an adequate Molecular Orbitals (MOs)
description of the  excited orbitals of the neutral molecule.
For $C_{\infty v}$ and $C_{2v}$,
the same symmetry subgroup is used in MOLPRO, namely $C_{2v}$,
the irreducible representations ($irrep$) of which
correspond to $\Sigma$ (and one $\Delta$),
$\Pi_x$, $\Pi_y,$ and $\Delta$ states, respectively, for the linear case, and
$A_1$, $B_2$, $B_1$, and $A_2$ states, respectively, for the isosceles case.
The MRCI was generated from a Complete Active Space (CAS) involving
three active electrons in 12, 8, 8, and 4
orbitals in each of the four  $irreps,$ respectively.
This means that  the generating CAS involves the valence orbitals
of H$_2$ as well as the 3$s$, 3$p$, 4$s$, 3$d$, 4$p$, 5$s$ orbitals of sodium
and even beyond. The MRCI space contains all simple and double
excitations with respect to the CAS space  (namely around 5~$\times$~10$^5$
configurations for each $irrep$).

\begin{table}  
  \caption{Calculated atomic transitions and errors from the sodium
    ground state,
    as compared to multiplet-averaged experimental data \citep{moore1971}
    (all in $cm^{-1}$)}
\begin{tabular}{c|c|c|c}
\hline
Level  &  present   &  exp &  $\Delta$\\
\hline
3p & 16944  & 16967    &  -23\\
4s & 25728  & 25740    &  -12\\
3d & 29159  & 29172    & -13\\
4p & 30250  & 30271    &  -21\\
5s & 33195  & 33200    & -5\\
\hline
\end{tabular}
\end{table}

Since we want to address spectral regions close to
the line center of the atomic absorbing lines 3$p$ and 4$p$, we
incorporated SO coupling
within a variant of the atom-in-molecule-like scheme introduced by 
\citet{cohen1974}.
This scheme relies
on a monoelectronic formulation of the spin-orbit coupling operator
\begin{equation}
  H_{SO}=\sum_i h_{SO}(i)=\sum_i\zeta_i \hat{l}_i . \hat{s}_i \; .
\end{equation}
The total Hamiltonian $H_{el}+H_{SO}$ is expressed in the basis set of the
 eigenstates (here with $M_s=\pm\frac{1}{2}$) of the purely electrostatic
 Hamiltonian $H_{el}$.
The spin-orbit coupling between the molecular  many-electron doublet
 states $\Phi_{k\sigma}$, approximated at this step as
single  determinants with the same  closed shell $\sigma_g^2$ $H_2$ subpart, 
is isomorphic to that
between  the  singly occupied molecular spin-orbitals $\phi_{k\sigma}$,
correlated with the six $p$  spin-orbitals of the alkali atom
($k$  labels the space part and $\sigma=\alpha,\beta$ labels the spin
projection).

 The Cohen and Schneider approximation  consists
in assigning these   matrix elements to their
asymptotic atomic values,
\begin{equation}
  <\Phi_{k\sigma}|H_{SO}|\Phi_{l\tau}>= 
  <\phi_{k\sigma}({\infty})|h_{SO}|\phi_{l\tau}({\infty})> 
.\end{equation}
The scheme
makes no  {\it a priori} assumption  about the magnitude of spin-orbit 
coupling versus pure electrostatic
interactions  and allows  general  intermediate coupling.
The main question for the applicability of the scheme in a basis of 
adiabatic states 
is the transferability of the atomic SO integrals to the molecular case,
because of configurational mixing. 
Such a situation 
characterizes the short-distance  interaction between the 
repulsive  state  correlated with
the $3p$ configuration (either 2$^2\Sigma^+$ 
or 2$^2A_1$, depending on the symmetry)  and the attractive $4s$
state (3$^2\Sigma^+$ or 3$^2A_1$).
This  means that the atomic spin-orbit coupling is not transferable at
short distance in the adiabatic basis.
Although this is not essential since the gap with the $^2\Pi$
components becomes large at 
short distance, we have taken  the variation of the
coupling scheme of the $3p,4s$ states  into account
in the following way. First, we achieved a diabatization of
the  $3p/4s$ anticrossing states in the 
$^2\Sigma^+$ (resp $^2A_1$) manifold, defining quasi-diabatic states 
$\tilde{\Phi}_k$ ($k=3p,4s$ omitting the spin mention for convenience
at this stage)
as states with a constant transition dipole moment 
from the ground state. The adiabatic states are related to the
latter through a 2x2 unitary transform at each distance $R$ depending
on a mixing angle $\theta$.
For the colinear case, the diabatic-to-adiabatic transformation is defined as 

\begin{eqnarray}
        \Phi(2^2\Sigma^+)=\cos\theta \tilde{\Phi}_{3p}(^2\Sigma^+)+\sin\theta\tilde{\Phi}_{4s}(^2\Sigma^+),\\
        \Phi(3^2\Sigma^+)=-\sin\theta \tilde{\Phi}_{3p}(^2\Sigma^+)+\cos\theta\tilde{\Phi}_{4s}(^2\Sigma^+)
,\end{eqnarray}
and  the same transformation holds in the  $C_{2v}$ symmetry for $^2A_1$ states.

Assuming the conservation of the transition
dipole moments from the ground state to the quasi-diabatic states along the
internuclear distance, the  transition moments to the adiabatic states can be
related to the former ones.
For instance, the transition moment from the ground state $\Phi_0$ to the
MRCI spin-orbit-less  eigenstate  $\Phi_{3p}$ is

\begin{eqnarray}
\mu(R)=<\Phi_0|z|\Phi_{3p}>=<\Phi_0|z|cos\theta\tilde{\Phi}_{3p}>=\cos\theta\times\mu^{at}_{3p} , 
\end{eqnarray}

where $\mu(R)$ is the MRCI spin-orbit-less
molecular adiabatic transition moment from the ground state to
eigenstate $\Phi_{3p}$
, and $\mu^{at}$ is its atomic or asymptotic value (2.537 $a_0^3$).
Such a transformation was only carried for distances at which the
molecular dipole moment is
less than its atomic value. In the medium range around
$R$=10-11 $a_0$, the adiabatic dipole transition moment 
 reaches a very shallow
maximum above its asymptotic value, that is  $2.565\;a_0^3$ for C$_{\infty v}$
and 2.549 $a_0^3$  for $C_{2v}$.  We note that complementarily,
the $\Phi_{4s}$ state,  asymtotically characterized by a
vanishing transition dipole moment, acquires
a non-zero coupling with the ground state with a $\sin\theta$ dependency.
The evolution of the mixing
along the internuclear distance is shown in Fig.~\ref{fig:diab}.
We obtain a  $8\times8$ spin-orbit coupling matrix for the 3$p$/4$s$ states, which is given in the Appendix.

The spin-orbit energy splitting  
$^2P_\frac{3}{2}-^2P_\frac{1}{2}$ of the 3$p$ levels of sodium is
17.19 $cm^{-1}$ (=$\frac{3}{2}\zeta$)
~\citep{moore1971}.
No such diabatization was considered for the $4p$ configuration
which has been treated 
via a 6x6 coupling matrix (e.g  $\cos\theta=1$) and a $\zeta$ constant
equal to one third of the  5.58 cm$^{-1}$
  experimental splitting of the 4$p$
manifold~\citep{moore1971}.
Spin-orbit coupling for the 3$d$ configuration has been  neglected.
 The diagonalization of the  total $H_{el}+H_{SO}$ matrix at each
 internuclear distance provides the final energies $E^{SO}_m$ and
 eigenstates $\Psi^{SO}_m$.

\begin{figure}
\centering
\vspace{8mm}
\includegraphics[width=8cm]{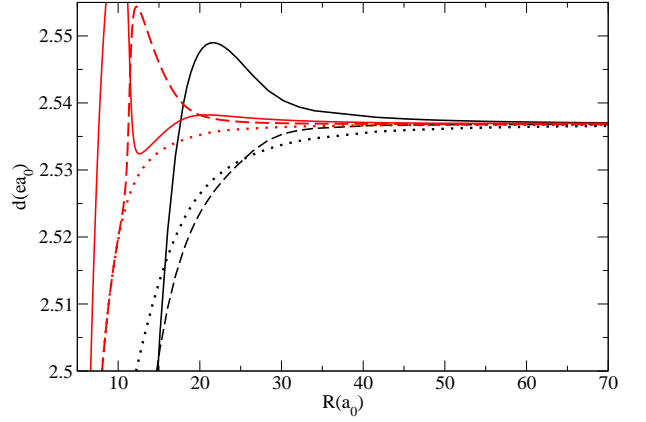}
\caption{Transition dipole moment  for the  $B$-$X$ (full line), 
$A$ $^2P_{3/2}$-$X$ (dotted line), and  $A$ $^2P_{1/2}$-$X$ (dashed line)
 transitions of the Na-H$_2$ molecule for the  C$_{\infty v}$  (red curves) and   
C$_{2v}$symmetries (black curves); $B$-$X$ (full line), 
$A$ $^2P_{3/2}-X$ (dotted line), 
and  $A$ $^2P_{1/2}-X$ (dashed line). }
\label{dip}
\end{figure}

The  transition  dipole moments shown in Fig.~\ref{dip} between the spin-orbit  
states were  determined by recombining  the 
adiabatic MRCI dipole moments over the coefficients $c^n_{{k\sigma}}$ of
the spin-orbit states $\Psi^{SO}_m$,
 \begin{equation}
\mathbf{D}^{SO}_{mn}=< \Psi^{SO}_m|\mathbf{D}|\Psi^{SO}_n>=\sum_{k\sigma,l\tau}c^m_{k\sigma}
c^n_{l\tau}<\Phi_{k\sigma}|\mathbf{D}|\Phi_{l\tau}>\delta_{\sigma\tau}.
\end{equation}

In the following section, we only focus on the states correlated with the 3$p$
manifold which determine the Na resonance lines.
We evaluate  the line parameters and collisional
profiles for relevant temperatures and
 densities that are appropriate for modeling  brown dwarf stars and hot-Jupiter-mass planets. 

\section{Temperature and density dependence of the Na resonance lines
  \label{sec:temperature}}

In \citet{allard1999}, 
we  derived a classical path expression for a pressure-broadened atomic
 spectral line shape that allows for an electric dipole moment that is
 dependent on the position of perturbers. 
This treatment has
 improved  the comparison of synthetic spectra of brown dwarfs with
 observations \citep{allard2003,allard2007b}.
 This approach to calculating the spectral line profile
requires knowledge of molecular  potentials with high accuracy because
the shape and  strength of the line profile are very sensitive to the details
 of the  molecular potential curves describing the Na--H$_2$ collisions. 
Sodium is the most abundant alkali in cool dwarf atmospheres and is mostly
responsible for the shape of  the optical spectrum. With precise
potentials and a complete line shape theory, 
major improvements in the theoretical description of pressure broadening
have been made compared to the commonly used van der Waals broadening in the
impact approximation \citep{burrows2003,allard2003,allard2012a,allard2012b}.

\subsection{Study of the line parameters \label{sec:parameter}}

The impact theories of pressure broadening 
\citep{baranger1958a,kolb1958} are based on the assumption of sudden
collisions (impacts)  between the radiator and
perturbing atoms, and are valid  when frequency displacements
\mbox {  $\Delta$ $\omega$ = $\omega$ - $\omega_0$} and gas densities are
sufficiently  small.

In impact broadening, the duration of the collision is assumed to be small
compared to the interval  between collisions, and the results describe the line
within a few line widths of center.
One outcome of our unified approach is that we may evaluate
the difference between the impact limit and the general unified profile, and
establish with certainty the region of validity of an assumed Lorentzian
profile.  
In the planetary and brown dwarf upper atmospheres 
the H$_2$ density is of the order of $10^{16}$ cm$^{-3}$
in the region of line core formation.

\begin{figure}
\resizebox{0.46\textwidth}{!}{\includegraphics*{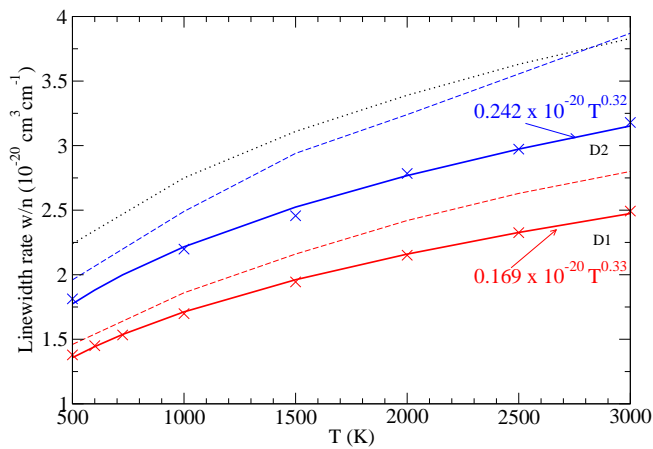}}
\caption{ Variation with temperature of the half-width
 of the $D$2 (blue curves) and $D$1  (red curves) lines of Na\,I perturbed
 by H$_2$ collisions. New {\it ab initio} potentials (full line), 
pseudo-potentials of ~\citet{rossi1985} (dashed lines), and the van der Waals 
potential (black dotted lines).
\label{width}}
\end{figure}

The line parameters presented in \citet{allard2007c},
\citet{allard2012a}, \citet{allard2012b} 
were obtained using the pseudo-potentials of \citet{rossi1985}.
To predict the impact parameters
the intermediate- and long-range part  of the potential energies 
need to be accurately determined. While the {\it ab initio} potentials presented in
\citet{allard2012b} allowed a better determination of the line wing, they
were not accurate enough to determine the line parameters.
With the improved potentials the full width at half-maximum $w$  is linearly 
dependent on H$_2$ density, and a power law in temperature is given  for the
 D1 line by
\begin{equation}
w = 0.169 \times 10^{-20} n_{\mathrm{H2}} \, T^{0.33},
\end{equation}
and for the D2 line is given by
\begin{equation}
  w = 0.242 \times 10^{-20} n_{\mathrm{H2}} \, T^{0.39}.
\end{equation}  
where $w$ is in cm$^{-1}$,   $n_{\mathrm{H2}}$ in cm$^{-3}$,  and $T$ in K.
These expressions accurately represent the numerical results as shown in Fig.~\ref{width},
and
may be used to compute the widths for temperatures of stellar or planetary
 atmospheres from 500 up to at least 3000~K.
 
\subsection{Line satellite}

Since the first Na--H$_2$ pseudo-potentials were obtained by \citet{rossi1985},
significant progress in the description of NaH$_2$
potentials has been achieved by \citet{burrows2003}, \citet{santra2005} and \citet{allard2012b}.
Blue satellite bands in alkali-He/H$_2$ profiles
can be predicted from the  maxima in 
the difference potentials $\Delta V$ for the $B$-$X$ transition.
Figures~1 and 2 of \citet{allard2012b} present the {\it ab initio} potential
curves without spin-orbit coupling for 
the  3$s$ and  3$p$ of S11 compared to pseudo-potentials of RP85.
It is seen there that the major difference with respect
 to S11 is that RP85 potentials are systematically less repulsive. This 
 difference     affects the blue satellite position.
The NaH$_2$ line satellite is  closer  to the main line than 
 obtained with RP85 (Fig.~5 of \citet{allard2012b}).
We observe this effect on synthetic spectra in the following section. 
 On the red side, the  NaH$_2$ wings match the profiles from the 
 RP85 potentials.

 \begin{figure}
\centering
\vspace{8mm}
\includegraphics[width=8cm]{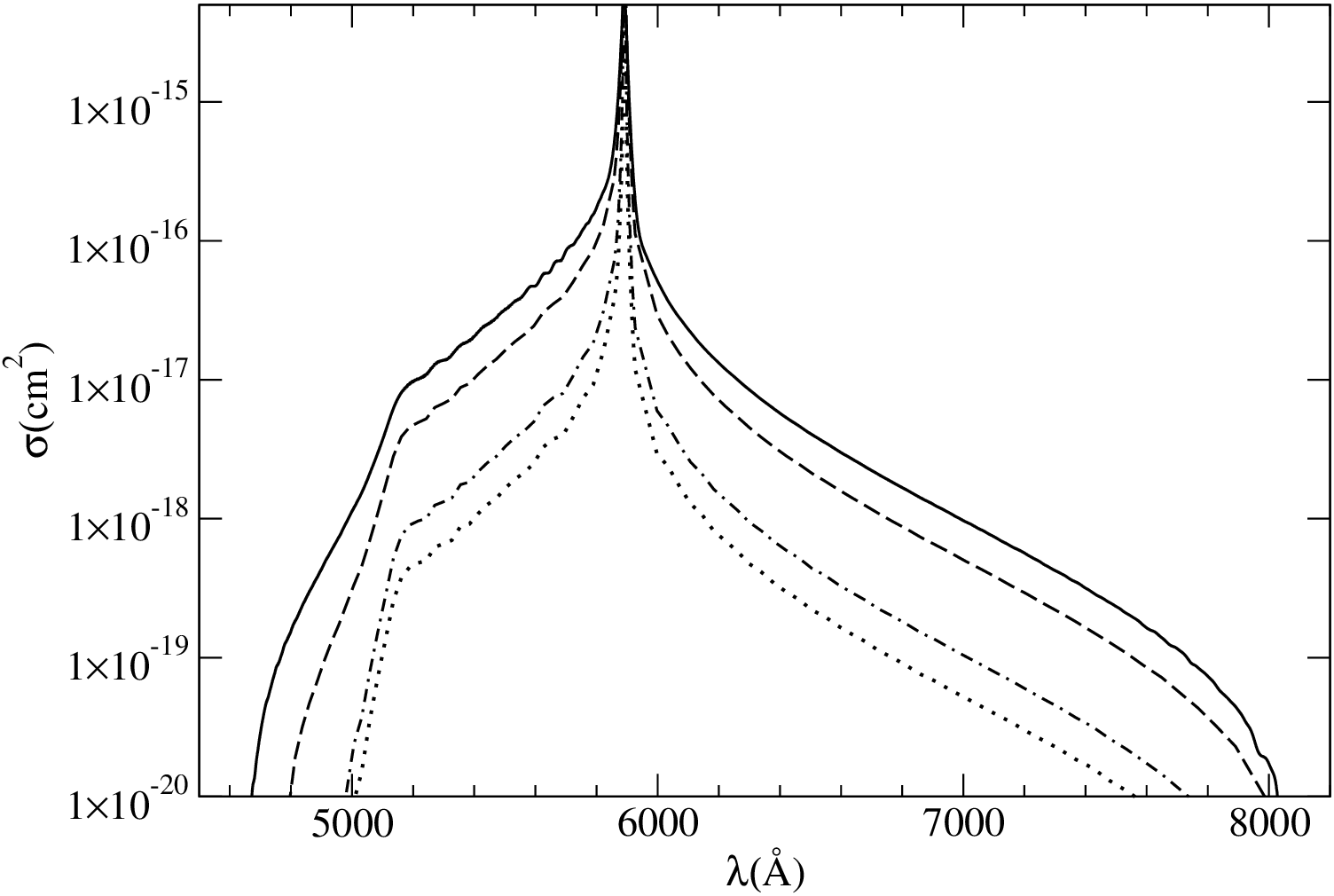}
\caption{Variation  with the  density   of   H$_2$  of the $D2$ 
  component
  (from top to bottom
  \mbox {$n_{\mathrm{H2}}$~=~$10^{21}$, 5$\times 10^{20}$, $10^{20}$
     and 5$\times 10^{19}$~cm$^{-3}$}).
  The  temperature is 1500~K.}
\label{D2vardens}
 \end{figure}
Figures~\ref{D2vardens}-\ref{D1varT} show the sensitivity of the
line wings to  pressure and temperature. 
The density effect on the shape of the blue wing  is  highly significant
 when the H$_2$ density becomes larger than 10$^{20}$  cm$^{-3}$.
We notice a first line satellite at 5170~\AA\ 
in Fig.~\ref{D2vardens}.
A second satellite due to multiple-perturber effects  appears as a shoulder at 
about  4800~\AA\/ for $n_{\mathrm{H2}} = 10^{21}$ cm$^{-3}$.
The density dependence of the far blue wing arises from multiple-perturber effects and is not  linear in density.
Figures~\ref{D2varT} and  \ref{D1varT} show the absorption
cross section of the resonance line of Na compared to the Lorentzian
profiles  calculated using the line widths 
presented in Fig.~\ref{width}, for T~=~1000~K.
The blue line
 wings shown in Fig.~\ref{D2varT} are almost unchanged with increasing 
 temperature whereas the red wings extend very far as temperature increases.
 \begin{figure}
  \resizebox{0.46\textwidth}{!}
            {\includegraphics*{35593fig8.eps}}
\caption{Variation of the absorption cross sections of the 3$s$-3$p$ D2 line
  component with temperature.
(from top to bottom  \mbox{$T$  =2500, 1500, 1000, and 600~K})  for $n_{\mathrm{H2}} = 10^{21}$ cm$^{-3}$. The Lorentzian profile for 1000~K is
  overplotted (black full line). 
 \label{D2varT}}
\end{figure}
\begin{figure}
  \resizebox{0.46\textwidth}{!}
            {\includegraphics*{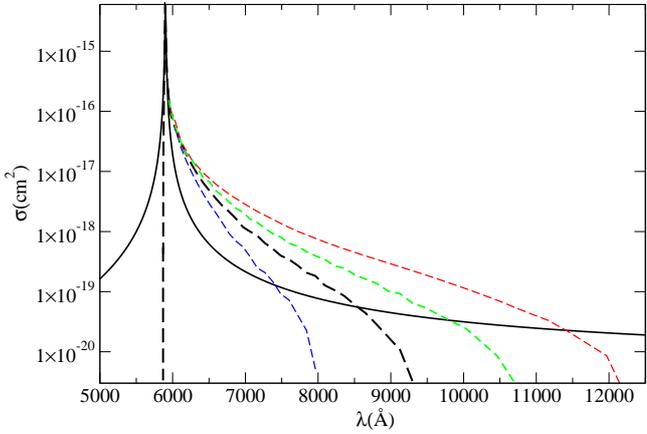}}
\caption  {Variation of the absorption cross sections of the 3s-3p D1 line
  component with temperature.
(from top to bottom  \mbox{$T$~=2500, 1500, 1000, and 600~K})
  for $n_{\mathrm{H2}} = 10^{21}$ cm$^{-3}$. The Lorentzian profile for 1000~K is
  overplotted (black full line).
\label{D1varT}}
\end{figure}

  \subsection{Opacity tables}
 For the  implementation of alkali lines perturbed by helium and molecular
hydrogen in atmosphere codes, the line opacity is calculated by splitting the
profile into a core component  described  with a Lorentzian profile, and
the line wings computed using an expansion of the autocorrelation function
in powers of density.
Here we  briefly  review the use of a density expansion in the opacity tables.

The spectrum $I(\Delta \omega)$ can be written as the Fourier transform (FT)
 of the dipole autocorrelation function $\Phi$(s) \citep{allard1999},
 
 \begin{equation}
I(\Delta \omega)=\frac{1}{\pi} \, Re \, 
\int^{+\infty}_0\Phi(s)e^{-i\Delta  \omega s} ds \; ,
\label{eq:int}
 \end{equation}
where $\Delta \omega$ is the angular frequency difference from the
 unperturbed center of the spectral line.
The autocorrelation function $\Phi$(s) is calculated with the
 assumptions that the radiator is stationary in space, the
perturbers are mutually independent, and in the adiabatic approach
 the interaction potentials give contributions that are scalar additive.
This last simplifying assumption allows us to calculate the total profile
 $I(\Delta \omega)$ when all the perturbers interact, as the
  FT of the $N^{th}$ power of
the autocorrelation function $\phi$(s) of a unique atom-perturber
pair. Therefore,
\begin{equation}
\Phi(s)=(\phi(s))^{N}\; ,
\end{equation}
 that is to say, we neglect the interperturber correlations.
We  obtain for a perturber density $n_p$ 
\begin{equation}
  \Phi(s) = e^{-n_{p}g(s)},
\label{eq:intg}  
\end{equation}
 where decay of the autocorrelation function with time leads to atomic line
 broadening.
When $n_p$ is high, the spectrum is evaluated by computing the
FT of Eq.~(\ref{eq:intg}).
The real part of $n_pg(s)$ damps
$\Phi(s)$ for large $s$ but this calculation is not  feasible when
extended wings have to be computed at low density  because of
the very slow decrease of the autocorrelation function.
An alternative is to use the
expansion of the spectrum  $I(\Delta \omega)$
in powers of the density described in \citet{royer1971a}.

We split the exponent $g(s)$ in  Eq.~(\ref{eq:intg})
 into a ``locally averaged part''
 $g_{{\rm av}}(s)$ and an ``oscillating
 part'' $g_{{\rm osc}}(s)$ by convolving $g(s)$ with a Gaussian $A(s)$:
 
 \begin{equation}
g_{{\rm av}}(s) =    A(s) * g(s),
\end{equation}
and
\begin{equation}
g_{{\rm osc}}(s) =  g(s) -g_{{\rm av}}(s),
\end{equation}
where the asterisk stands for a convolution product.\\
We can write 
\begin{displaymath}
 g(s)  = g_{{\rm av}}(s) + g_{{\rm osc}}(s).
\end{displaymath}

At large values of $s$, $g(s)$ becomes linear in $s$, $g_{{\rm osc}}$ vanishes,
 and the oscillating part remains bounded which allows us to expand
  $e^{-n_{p}g_{{\rm osc}}(s)}$ in powers of $n_{p}g_{{\rm osc}}(s)$; 
Eq.~(\ref{eq:intg}) becomes
\begin{equation}
\Phi(s)  =  e^{-n_{p}g_{{\rm av}}(s)}[ 1 - n_{p}g_{{\rm osc}}(s)
  + \frac{n_{p}^2}{2!}[g_{{\rm osc}}(s)]^2 + \ldots].
\label{eq:gexp}
\end{equation}
The complete profile is given by the FT of Eq.~(\ref{eq:gexp}):
\begin{equation}
  I(\Delta \omega) = I_c(\Delta\omega)*[\delta(\Delta \omega) -
    n_{p}I_w(\Delta \omega)
   + \frac{n_{p}^2}{2!}[I_w(\Delta \omega)]^{*2} + \ldots]
\label{eq:Iexp}
,\end{equation}
where 
$I_c(\Delta \omega)$ = FT[$e^{-n_{p}g_{{\rm av}}(s)}$] forms the core of the
 line profile and  $I_w(\Delta \omega)$ = FT[$g_{{\rm osc}}(s)$] is
responsible  for the wing.

This method gives the same results as the FT of the general
autocorrelation function (Eq.~(\ref{eq:intg}) without density expansion) 
at higher densities
and has the advantage of including  multiperturber effects at very low
density when the general calculation is not feasible
\citep[see, e.g.,][]{allard2013c}.
The impact approximation determines the asymptotic 
behavior of the  unified line shape correlation function.  In this way
the results described here are applicable to a more general
line profile and opacity evaluation for the same perturbers 
at any given layer in the photosphere or planetary atmosphere.

When the 
expansion is stopped at the first order it is equivalent to the
one-perturber approximation. Previous opacity tables were constructed to third order allowing us to obtain line
profiles up to $N_{H2}$=$10^{19}$ cm$^{-3}$. The new tables  are constructed
to a higher order allowing line profiles to $N_{H2}$=$10^{21}$ cm$^{-3}$.

 For a more  direct comparison of the contributions of the
two fine-structure components of the doublet  
it is convenient to use a cross-section $\sigma$ associated to each
component.
The relationship between the computed cross-section and the normalized absorption coefficient given in 
Eq.~(\ref{eq:int}) is
\begin{equation}
  I(\Delta \omega) = \sigma(\Delta \omega) / \pi r_0 f \; ,
\label{eq:sig}  
\end{equation}
where $r_0$ is the classical radius of
the electron, and $f$ is the oscillator strength of the transition.

\begin{figure*}
  \begin{center}
\includegraphics[width=1.00\textwidth]{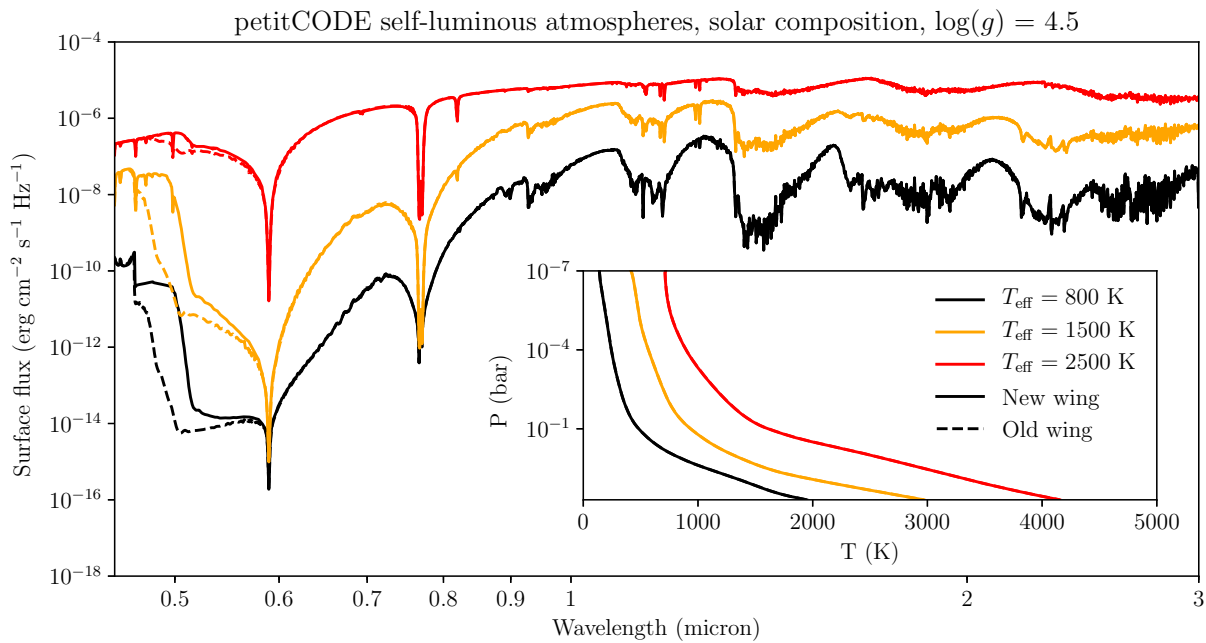}
\caption{Emission spectra for cloud-free, self-luminous
   objects (exoplanets or Brown Dwarfs) at
  solar composition and varying effective temperature, calculated
  with {\it petitCODE} \citep{mollierevanboekel2015,mollierevanboekel2016}.
  The atmospheric surface gravity was set to ${\rm log}(g)=4.5$,
  with $g$ in units of $\rm cm \ s^{-2}$. The black, orange, and red lines
  show the spectra for planets with $T_{\rm eff}$ = 800, 1500, and 2500 K,
  respectively. Solid lines denote results obtained with the new Na wing
  profiles (presented in this paper), whereas dashed lines denote the
  results obtained with the Na lines reported in \citet{allard2003}.
  The {\it inset plot} shows the self-consistent temperature profiles
  for these cases, as calculated with  {\it petitCODE}. In the example shown
  here the effect of the change in opacity of the Na wings is too small
  to be seen.}
\label{fig:comp_bd}
\end{center}
\end{figure*}

\begin{figure*}  
\begin{center}
\includegraphics[width=1.\textwidth]{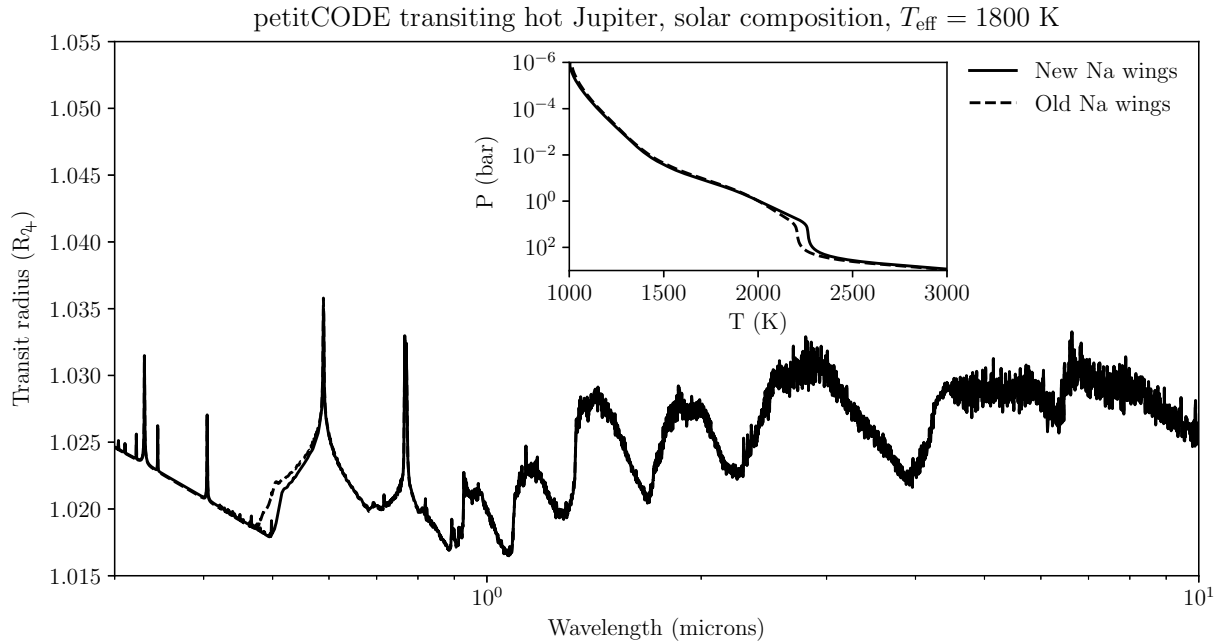}  
\caption{Transmission radii for cloud-free, hot-Jupiter exoplanets at solar composition for a planetary effective temperature of 1800 K, calculated with
  {\it petitCODE} \citep{mollierevanboekel2015,mollierevanboekel2016}.
  The planet mass and radius were chosen to be identical to the values
  of Jupiter, and an internal temperature of $T_{\rm int}=200$~K was used.
   The opacity of TiO and VO was neglected.
  The host star was chosen to be a solar twin. Solid lines denote results
  obtained with the new Na wing profiles (presented in this paper),
  whereas dashed lines denote the results obtained with the Na lines
  reported in \citet{allard2003}. The {\it inset plot} shows the
  self-consistent temperature profiles for these cases, as calculated
  with  {\it petitCODE}. In the example shown here the weaker absorption
  in the new Na wing profiles leads to more greenhouse heating in the
  deep layers of the atmosphere, while the upper layers are slightly
  cooler than what was obtained with the old profiles.}
\label{fig:comp_trans}
\end{center}
\end{figure*}
 
\section{Astrophysical applications \label{sec:astro}}

\subsection{Self-luminous atmosphere}

 In Fig. \ref{fig:comp_bd} we show the emission spectra for a cloud-free,
 self-luminous object (exoplanet or brown dwarf)
 at solar composition and varying effective
temperature, calculated with {\it petitCODE}
\citep{mollierevanboekel2015,mollierevanboekel2016}.
The atmospheric surface gravity was set to ${\rm log}(g)=4.5$,
with $g$ in units of $\rm cm \ s^{-2}$. The effective temperature
was set to $T_{\rm eff}$ = 800, 1500, and 2500 K, respectively.
We calculated atmospheric structures and spectra using the old and new
Na--H$_2$ line profiles.  The opacity of TiO, VO, and FeH was neglected to
 make the alkali lines visible also for the highest-temperature model.
The difference in Na blue wing absorption is
clearly visible.
We also show the self-consistent temperature profiles of the atmospheres
for these cases, as calculated with  {\it petitCODE}. In the example shown
here the effect of the change in opacity of the Na wings on the
 temperature profile is too small to
be seen and the solid and dashed lines overlap.

 \subsection{Hot Jupiter}
 In Fig.~\ref{fig:comp_trans} we show the transmission spectra for
 cloud-free hot-Jupiter exoplanets at solar composition for a planetary
 effective temperature of 1800 K, also calculated with {\it petitCODE}
 \citep{mollierevanboekel2015,mollierevanboekel2016}. The planet mass and
 radius were chosen to be identical to the values of Jupiter, and an
 internal temperature of $T_{\rm int}=200$~K was used.
  The  TiO/VO opacities were  neglected.
 The host star was
 chosen to be a solar twin. We calculated atmospheric structures and spectra
 using the old and new NaH$_2$ line profiles.
 The difference in Na blue wing absorption
 is clearly visible. We also show the self-consistent temperature profiles
 of the atmospheres for these cases, as calculated with {\it petitCODE}.
 In the example shown here the weaker absorption in the new line profiles
 leads to more green house heating in the deep layers of the atmosphere,
 while the upper layers are slightly cooler than what was obtained with
 the old line profiles.

\section{Conclusion}
  We performed  theoretical  calculations of the collisional profiles 
of the resonance lines of Na  perturbed by H$_2$
  using a unified theory of spectral line
  broadening and high-quality  {\it ab initio} potentials and transition
  moments.
Figures~\ref{fig:comp_bd} and \ref{fig:comp_trans} show that the perturbation of
Na by H$_2$  can be  very important for  the interpretation of visible
spectra of brown dwarf and exoplanet atmospheres.   We therefore suggest that  the use of Lorentzian
profiles is  not appropriate for modeling  the line wings, as 
Figs.~\ref{D2varT} and  \ref{D1varT} clearly show.  Complete unified line profiles
based on accurate
atomic and molecular physics should be incorporated into analyses of exoplanet
spectra when precise absorption coefficients are needed.
Calculations are presented for the $D1$ and $D2$ lines
 from $T_\mathrm{eff}=500\ \mathrm{K}$ to 3000 K with a step size of
 500 K.  Tables of the density expansion coefficients,
   an explanation of their use, and a program to produce
   line profiles to $N_{H2}$=$10^{21}$ cm$^{-3}$
   will be archived at the CDS.


\appendix
\section{SO matrix for $3p$ and $4s$ states}

 \begin{center} 
\begin{displaymath}
\left(
\begin{array}{cccccccc}
E_{3px} &0 &-i\frac{\zeta}{2}  &0 &0 &\frac{\zeta}{2}cos\theta&0&-\frac{\zeta}{2}sin\theta  \\
0 &E_{3px} & 0 &i\frac{\zeta}{2} & \frac{\zeta}{2}cos\theta &0&-\frac{\zeta}{2}sin\theta&0  \\
i\frac{\zeta}{2}& 0 & E_{3py} &0 &0 &-i\frac{\zeta}{2}cos\theta&0&i\frac{\zeta}{2}sin\theta  \\
0 &-i\frac{\zeta}{2} & 0 &E_{3py} &-i\frac{\zeta}{2}cos\theta &0&i\frac{\zeta}{2}sin\theta&0  \\ 
0&\frac{\zeta}{2}cos\theta & 0 &i\frac{\zeta}{2}cos\theta &E_{3pz}&0&0 & 0 \\
\frac{\zeta}{2}cos\theta &0 & i\frac{\zeta}{2}cos\theta &0& 0&E_{3pz}&0&0 \\
0&-\frac{\zeta}{2}sin\theta&0&-i\frac{\zeta}{2}sin\theta&0&0&E_{4s}&0\\
-\frac{\zeta}{2}sin\theta&0&-i\frac{\zeta}{2}sin\theta&0&0&0&0&E_{4s}
\end{array}
\right)
\end{displaymath}
\end{center}

\end{document}